\documentclass[]{article}
\usepackage{authblk}
\usepackage{graphicx}
\usepackage{amsmath}
\usepackage{amsfonts}
\usepackage{enumitem}

\title{Characterization of a vertical crack by means of local thermal analysis}

\author[1]{Gabriele Inglese}

\author[2]{Roberto Olmi}

\author[1]{Agnese Scalbi}

\affil[1]{IAC ``M. Picone'' - CNR, Via Madonna del Piano 10, 50019 Sesto Fiorentino, Italy}
\affil[2]{IFAC  - CNR, Via Madonna del Piano 10, 50019 Sesto Fiorentino, Italy}

\begin{document}

\date{}

\maketitle

\begin{abstract}
This paper deals with the solution of an inverse problem for the heat equation aimed at nondestructive evaluation of fractures. A fundamental step in any typical iterative inversion method, is the numerical solution of the underlying direct mathematical model. Usually, this step requires specific techniques in order to limit an abnormal use of memory resources and computing time due to excessively fine meshes necessary to follow a very thin fracture in the domain. Our contribution to this problem consists in decomposing the temperature of the damaged specimen is the sum of a term  (whose analytical form is known) due to an infinite virtual  fracture plus the solution of an initial boundary value problem for the heat equation on one side of the fracture (i.e. on a rectangular domain). The depth of the fracture is a variable parameter in the boundary conditions that must be estimated from additional data (usually, measurements of the surface temperature). We apply our method to the detection of simulated cracks in concrete and steel specimens. 
\end{abstract}

\section{Introduction}

The ideas behind the present note arise in the context of thermographic analysis of fractures on the surface  of reinforced concrete artifacts \cite{IOS18} or metal plates \cite{POMOCS14bis}. In particular, we deal with the mathematical modeling of Laser Spot Thermography (LST) (\cite{GIFK83} is one of the oldest paper about laser and cracks while   \cite{IAKN19} is one of the most recent items in a huge list). LST  is a special case of Active Thermography \cite{Ma01} in which high optical intensities are applied to a small spot very close to an emerging fracture.
In order to obtain a non destructive evaluation of the fractures properties, we must derive  approximate solutions of an inverse problem for the heat equation. A very recent overview about thermal imaging of defects, including a large and updated references section,  is in \cite{DWW19}.\\

Though the problem is posed  naturally in a parallelepipedal slab  $(-L,L) \times (-L,L) \times (0,a)$, we consider the heat equation in the half-space obtained for $L,a \to \infty$. Time is in the interval $(0,t_{max}]$. Actually, we do not lose generality as long as both $L$ and $a$ are large enough to be considered infinite in the time scale we are dealing with. Numerical examples about temperature increase, diffusivity and time scale, in the case of different materials, are reported in section \ref{sec:diffusivity}. For the same reasons the inverse analysis of the stationary problem is, most of the time, meaningless.\\

In this paper we deal with 2D models. We recall that a 2D continuous point source close to the emergent point of a linear crack is equivalent to a 3D continuous line  source parallel (at a distance $x_0$) to a plane fracture of length  $\lambda >> x_0$ in a 3D slab. \\

Let the half plane $\Omega=\{ x \in (-\infty,\infty),\phantom{a}z\in (0,\infty)\}$ represent a 2D slab whose face $z=0$ is heated by a Laser Spot centered in the origin $(0,0)$. Laser is kept on for a time $t_{laser}<t_{max}$ resulting in a temperature $u^0>0$ in $\Omega \times (0,t_{max}]$ (background temperature) whose exact analytical form is known (see section \ref{sec:back}).\\

Let $\sigma$ be the segment $\{(x_0,z)\}_{0<z<b\le \infty}$ with fixed $x_0>0$ (one dimensional linear vertical crack) . The same laser source as before, in presence of the defect $\sigma$,  determines the temperature  $u^b(x,z,t)$ in $\Omega \setminus \sigma \times (0,t_{max}]$.  The thermal conductance of $\sigma$ is $H \ge 0$.

We reasonably assume that a fracture evolves in a time scale much greater than $t_{max}$. Hence, it is not restrictive to consider $\sigma$ constant in $t$.\\

In order to establish a quantitative relation between the temperature $u^b(x,0,t)$ observed at the surface of the specimen and the depth of the crack $\sigma$, we write the restriction of $u^b(x_0+\xi,z,t)$ to the quarter $\xi > 0$
(i.e. on the right hand side of the crack) as $u^b(x_0+\xi,z,t)=C_H(\xi,z,t)+E_b(\xi,z,t)$, 
where $C_H$ is the temperature on the right hand side of an infinite  crack with thermal conductance $H$ (its analytical form is derived in section \ref{sec:infinitecrack} following \cite{CJ59} and \cite{POMOCS14})
and $E_b$ is the solution of a well posed Initial Boundary Value Problem (IBVP) for the heat equation (see section \ref{sec:Ebi}) in a rectangular domain. \\

Despite of the fact that no analytical expression of the temperature in presence of a finite crack is available \cite{POMOCS14bis}, the expression $C_H(\xi,0,t)+E_b(\xi,0,t)$ for the surface temperature makes the evaluation of the depth of the crack a simpler task than using discontinuous finite elements in a standard optimization procedure.  In section \ref{sec:inverse} we use the full knowledge of $C_H$ and $E_{b}$  to identify the unknown depth of the fracture when surface measurements (or simulations, in our case) of $u^b(x,0,t)$ are available.\\

The assumption that an ideal crack is a vertical segment is justified by the shape of a wide class of real fractures \cite{J96} \cite{B02}. In fact, more than the exact shape of the fracture, it is interesting to evaluate average parameters like width, length and depth.  In section \ref{sec:inverse} we simulate the collection of temperature maps of the surface temperature in presence of unknown cracks which deviates from the ideal linear geometry. Then, we compute, by means of the method developed in this paper, a kind of order zero approximation that works well in the evaluation of the depth of the fracture. 

\section{Geometry, definitions and Laser Spot Thermography}

A vertical {\it  ideal crack} is a segment orthogonal to the lower face of a parallelepiped-shaped specimen.
Although the geometrical figure
that represents a slab  is clearly the rectangle $\Omega_{a,R} = (-R,R)  \times (0,a)$ with $0< a < R$, we assume  in what follows that our specimen is the half-plane $\Omega = \{z \ge 0\}$. The ideal crack is $\sigma = \{(x_0,z)\}_{0<z< b \le \infty}$. 
Since we deal with relatively short-lasting measurements (real or simulated) these assumptions are not restrictive (see section \ref{sec:diffusivity}).
\\

A {\it physical crack} (or {\it fracture}) $C_\epsilon$ is an open neighborhood of $\sigma$. We suppose here that the set $C_\epsilon$ is the rectangle $(x_0-\epsilon,x_0+\epsilon) \times (0,b)$.

Though a real fracture has hardly a perfectly rectangular section,
 this simplification is useful to characterize unknown fractures. Indeed, vertical cracks
 has been recently  studied in  \cite{POMOCS14} \cite{POMS16} \cite{BS17} \cite{IOS18}.\\

Finally, we define the domains
\begin{equation}
\Omega_0 = \Omega \setminus  \sigma
\end{equation}
and
\begin{equation}
\Omega_\epsilon = \Omega \setminus  C_\epsilon.
\end{equation}

\subsection{Time scales, diffusivity and characteristic dimensions of the slab}\label{sec:diffusivity}

In this section, we compare the thermal behavior of the  slab of thickness $a$ with the  half-plane obtained for $a \to \infty$.   Assume that, in both cases, our specimen is heated by the constant flux $\frac{\phi_0}{\kappa}$ applied to the  surface $z=0$. The corresponding temperatures $u_a$ and $u_\infty$ are
\begin{equation}
u_a(z,t) = \frac{2\phi_0\sqrt{\alpha t}}{\kappa} \sum_{n=0}^\infty \left( \mathrm{ierfc}\frac{2 n a + z}{2\sqrt{\alpha t}} + \mathrm{ierfc}\frac{2 (n+1) a - z}{2\sqrt{\alpha t}}  \right)
\label{eq:slab}
\end{equation}
(\cite{CJ59} pag 112)
and its limit for $a \to \infty$
\begin{equation}
u_\infty(z,t) = \frac{2\phi_0}{\kappa} \left[ \sqrt{\frac{\alpha t}{\pi}} e^{-\frac{z^2}{4\alpha t}} -\frac{z}{2} \mathrm{erfc}\left(\frac{z}{2\sqrt{\alpha t}} \right) \right]
\label{eq:inf}
\end{equation}
(directly derived in \cite{CJ59} pag 75).\\

As long as   $u_a(0,t)$ and $u_\infty(0,t)$  coincides within acceptable error limits, we say that, for our purposes,  the semi-infinite medium is a useful approximation of the slab. \\

In what follows, we give an estimate of the relative deviation
 $D_a(t) = \frac{u_a^1(0,t) - u_\infty(0,t)]}{u_\infty(0,t)} $  in the time interval  $t \le \tau_{ch}=\frac{a^2}{\alpha}$ where $\tau_{ch}$  is a {\it characteristic time } in the sense that diffusion through a distance $a$ takes roughly $\tau_{ch}$ time units (see for example \cite{LS74}).
\\
  
We evaluate  (\ref{eq:slab}) and (\ref{eq:inf}) for $z=0$ and obtain
\begin{equation}
u_a(0,t) = \frac{2\phi_0\sqrt{\alpha t}}{\kappa} \sum_{n=0}^\infty \left( \mathrm{ierfc}\frac{n a}{2\sqrt{\alpha t}} + \mathrm{ierfc}\frac{(n+1) a}{2\sqrt{\alpha t}} \right)
\label{eq:slab_a}
\end{equation}
and
\begin{equation}
u_\infty(0,t) = \frac{2\phi_0 \sqrt{\alpha t}}{\kappa \sqrt{\pi}} 
\label{eq:inf_a}
\end{equation}
where
\begin{equation}
\mathrm{ierfc}(x) = \frac{e^{-x^2}}{\sqrt{\pi}} - x\  \mathrm{erfc}(x).
\end{equation}
and erfc(x) is the complementary error function.

It can be verified numerically that the first few terms of the series in  (\ref{eq:slab_a}) are sufficient to compute the temperature $u_a$. Actually, the first term
\begin{equation}
u_a^1(0,t) = \frac{2\phi_0 \sqrt{\alpha t}}{\kappa} \left[ \frac{1}{\sqrt{\pi}}  + \mathrm{ierfc}\left(\frac{a}{\sqrt{\alpha t}} \right) \right] 
\end{equation}
is a already a good approximation (very good, for low-conducting materials).   If we write the relative deviation $D_a(t)$
in terms of the adimensional variable $\xi = \sqrt{\frac{\tau_{ch}}{t}}$ we obtain
\begin{equation}
D_a(t) =\sqrt{\pi} \mathrm{ierfc}(\xi) = e^{-\xi^2}-\sqrt{\pi}\xi  \mathrm{erfc}(\xi).
\end{equation}

Taking the asymptotic expansion
\begin{equation}
\mathrm{erfc(\xi)} \approx \frac{e^{-\xi^2}}{\xi \sqrt{\pi}} \left( 1 - \frac{1}{2\xi^2}) \right)
\end{equation}
valid  for relatively large values  of $\xi$,
we finally obtain
\begin{equation}
D_a(t) \approx \frac{1}{2} \frac{e^{-\xi^2}}{\xi^2}.
\label{eq:approx_delta}
\end{equation}
so that the relative deviation approaches zero as fast as $e^{-\frac{\tau_{ch}}{ t}}$ for $t \to 0$. 

For example, if $\xi\ge 2.1$ (i.e. $t \le \frac{\tau_c}{4.41} $) we obtain $D_a( t) \le .03 $ (i.e. we have a relative deviation lower than $3\%$).\\

We verify the effectiveness of the semi-infinite approximation of our slab for two different materials: a slab of concrete ($\kappa = 1$, $\rho = 1900$, $c = 1000$, all in MKS units)
with a thickness $\mathrm{a = 4\ cm}$, a slab of stainless steel ($\kappa = 14.4$, $\rho = 8000$, $c = 500$, all in MKS units) with a thickness $\mathrm{a = 2\ cm}$. We obtain (the headers ``c'' and ``s'' respectively denote ``concrete'' and ``steel''):

\begin{align}
& \tau_{ch}^c(a = 4 cm) \approx 1400 s\\
& \tau_{ch}^s(a = 2 cm) \approx 50 s
\end{align}

In other words, if we use $u_\infty$ (i.e. a much simpler function) instead of $u_a$ we introduce a negligible error in our procedure up to about 1400 s (characteristic time) for a concrete wall of 4 cm thickness.  For a stainless steel slab of 2 cm thickness, the characteristic time is about 50 seconds.\\

\begin{figure}[!h]
\centering
\includegraphics[width=10cm]{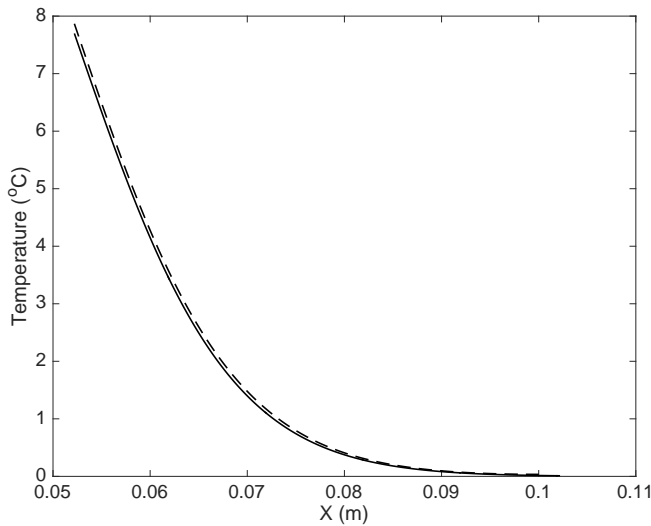}
\caption{Comparison between the ``true'' (dashed line) and analytical (solid line) temperature (see text)}
\label{confr}
\end{figure}

In Figure \ref{confr}  we compare the ``true'' temperature (dashed line) on the right hand side of a passing-through crack in a finite specimen of 20 mm thickness, obtained by a FEM simulation, with the function $u_\infty(x,0,t)$  (solid line). Temperatures refer to time t = 40 s.

\subsection{Laser spot on the accessible surface of a slab: Background temperature} \label{sec:back}

In Laser Spot Thermography, a specimen  at room temperature  $U^0$ is heated  by means of a laser. When the specimen is undamaged, the temperature increase $u^0= T-U^0$, due to the application of the laser source, fulfills  the heat conduction equation (T is the absolute temperature)
\begin{equation}
\rho c u_t(x,z,t) = \kappa \Delta u(x,z,t)
\end{equation} 
with $(x,z) \in \Omega$  
and $t \in(0,t_{\max}]$ provided that the following
boundary conditions are satisfied
\begin{equation}
-\kappa u_z(x,0,t)=
\phi(x,t)
\end{equation}
\begin{equation}
lim_{x^2+z^2 \to \infty} u^2 =0.
\end{equation}
The physical parameters $\rho$, $c$ and $\kappa$ are, respectively,  density, specific heat and thermal conductivity of the material under investigation while $\alpha=\frac{\kappa}{\rho c}$ is the diffusivity.

Since $\chi_E(x)=1$ if $x \in E$ and $\chi_E(x)=0$ elsewhere, the laser spot has in theory the form $\phi_L\delta(x)\chi_{(0,t_L)}(t)$ (i.e. the laser is ON from $t=0$ to $t=t_L<t_{\max}$). This is called a continuous point source (in 2D). If regarded in a 3D framework this is a continuous line source, constant in the direction $y$. In practical models and simulations, laser spot has the rectangular shape
\begin{equation}
\phi(x,t) = \phi_L \chi_{(-\delta_L,\delta_L)}(x)\chi_{(0,t_L)}(t).
\label{power}
\end{equation}
The constant $\phi_L>0$ is the maximum power per unit surface of the laser source,  while the ``deviation''  parameter $\delta_L>0$ (power width, in the following) measures the spread of the Laser Spot. 

Since we deal with relative temperatures, the initial value is:
\begin{equation}
u(x,z,0)= 0
\end{equation}
in $\Omega$.\\

The solution of the above IBVP for the undamaged specimen is called ``background temperature''. It is denoted by $u^0$ and it  can be easily obtained in terms of the Green's function for the half-plane \cite{CJ59}:

\begin{equation}
u(r,t) = \frac{\phi_L}{2\pi\kappa} \int_{-\delta_L}^{\delta_L} \left[ -\mathrm{Ei}\left( -\frac{r^2}{4\alpha t} \right) + \mathrm{Ei}\left( -\frac{r^2}{4\alpha (t - t_L)} \right) \right] dx'
\label{line_source}
\end{equation}
with $r^2 = (x-x')^2+z^2$ and $t > t_L$. For $t < t_L$ the second right hand side of (\ref{line_source}) is zero. $Ei$ is the exponential-integral function.

\subsection{Laser spot on the accessible surface of a slab: Temperature of the damaged slab} \label{sec:dama}
 
A  physical crack  $C_\epsilon$ is supposed to have  a positive  conductivity $\kappa_c$. Since the width of $C_\epsilon$ is $2\epsilon$, the thermal conductance  of the crack is  $H \approx \frac{\kappa_c}{2\epsilon}$ ($R_{th}=\frac{1}{H}$ is called thermal resistance)(see \cite{CJ59} sect 1.9).  When the heat conduction through the fracture is actually negligible, we say that the crack is insulating and $H=0$. \\

The model of heat conduction in the fractured slab consists of a system of two parabolic equations
 \begin{equation}\label{equequ0}
\rho   c   u_t(x,z,t) = \kappa   \Delta u(x,z,t)
\end{equation} 
with $(x,z) \in \Omega_\epsilon$ and $t \in(0,t_{\max}]$
and
\begin{equation}
\rho_c c_c u_t(x,z,t) = \kappa_c \Delta u(x,z,t)
\end{equation} 
with $(x,z) \in C_\epsilon$ and $t \in(0,t_{\max}]$ (the subscript $c$ means ``crack'').

We assume that $0<\delta_L<x_0-\epsilon$ (i.e. no heating is revealed on the right of the crack). Here, the unknown function $u$ is the temperature increase $u=T-U^0$ where $T$ is the absolute temperature of the damaged slab and $U^0$ is the initial and room temperature. The boundary condition for $z=0$ is
\begin{equation} \label{equequ1}
-\kappa   u_z(x,0,t)=\phi(x,t).
\end{equation}

The temperature increase  vanishes at infinity and
its initial value is
\begin{equation} \label{equinitial}
u(x,z,0)= 0
\end{equation}
in $\Omega$.\\

Let $u^\epsilon  $ and $u^\epsilon_c$ fulfill the IBVP, respectively in $\Omega_\epsilon$ and $C_\epsilon$, for all times $t$.  Here we assume for simplicity that $C_\epsilon$ is the thin rectangle $(x_0-\epsilon,x_0+\epsilon)\times(0,b)$. At the interfaces $\{ x = x_0 \pm \epsilon, \  z \in (0,b)\}$ we have transmission conditions  for  temperature 
$$u^\epsilon  (x,z,t)=u^\epsilon_c(x,z,t)$$
 and heat flux 
\begin{equation}\label{IBVP01}
\kappa   u^\epsilon_{x}(x,z,t)=\kappa_c u^\epsilon_{cx}(x,z,t)
\end{equation}
 where the subscript $x$ means partial derivative of $u^\epsilon  $ and $u^\epsilon_c$.  \\

We know from \cite{IOS18} that 
\begin{equation}
\lim_{\epsilon \to 0} u^\epsilon = u^b
\end{equation}
where $u^b$ is the temperature of the ideal cracked domain $\Omega \setminus \sigma$.  In the limit for $\epsilon \to 0$, transmission conditions become   
Robin boundary conditions on the two sides of $\sigma$
\begin{equation}\label{bcleft}
\kappa   u_{b x}(x_0,z,t)+2H(u^b(x_0^-,z,t)-u^0(x_0,z,t))=0
\end{equation}
\begin{equation}\label{bcright}
-\kappa    u_{b x}(x_0,z,t)+2H(u^b(x_0^+,z,t)-u^0(x_0,z,t))=0
\end{equation}
where $u^b(x_0^\pm,z,t) = \lim_{\epsilon \to 0} u^\epsilon(x_0\pm \epsilon,z,t)$  and
\begin{equation}
H = \lim_{\epsilon \to 0}  \frac{\kappa_c}{2\epsilon}
\end{equation}

The evaluation of $H$ is an inverse problem in itself when the fracture is actually an interface between different materials \cite{GBMS19}.

 \subsection{Humps theorem}
 
 Using the notation of previous subsection:\\
 
 \noindent {\it Theorem} For all $b>0$,  $t \in (0,t_{\max}]$ , $\xi \in (0,\infty)$, $z \ge 0$ and $x_0>0$ we have
\begin{equation}\label{invar}
u^b(x_0+\xi,z,t)+u^b(x_0-\xi,z,t)=u^0(x_0+\xi,z,t)+u^0(x_0-\xi,z,t).
\end{equation}

\noindent {\it Proof} We obtain the following symmetric  IBVP for heat equation by reflection (with respect to the axis $x=x_0$)  of the problem with a crack $\sigma$:
\begin{equation}\label{equ00sym}
\rho   c   v_t(x,z,t) = \kappa   \Delta v(x,z,t)
\end{equation} 
where $\sigma=\{(x_0,z)\}_{z \in(0,b)}$ and  $(x,z) \in \Omega_\sigma = \Omega \setminus \sigma$ and $t \in(0,t_{\max}]$. The boundary condition on the surface of the slab is
\begin{equation}
-\kappa   v_z(x,0,t) + H v(x,0,t) =  
\phi(x,t)+\phi(x-2 x_0,t)
\end{equation}
where $\phi$ is defined in (\ref{power}). The gradient of $v$ vanishes at infinity and
the initial temperature is still
\begin{equation}\label{equ01sym}
v(x,z,0)= 0
\end{equation}
in $\Omega$.

The solution of (\ref{equ00sym})-(\ref{equ01sym}) is $v(x,z,t)=u^b(x_0+\xi,z,t)+u^b(x_0-\xi,z,t)$ for linearity. Moreover, $v$  is smooth and symmetric with respect to $x=x_0$. It means that  $v_x(x_0,z,t)=0$ $\forall t$ and $z$ i.e. the problem on the half-plane
$x>x_0$ (or $x<x_0$) has a unique solution that does not depend on $b$. \\ \\

\noindent {\it Corollaries}

(i) $u^b_x(x_0^+,z,t)=u^b_x(x_0^-,z,t)$ for all $z$ and $t$ (in particular, for $z \in (0,b)$ when $u$ is not continuous)

(ii) $\frac{u^b(x_0^+,z,t)+u^b(x_0^-,z,t)}{2}=u^0(x_0)$\\ \\

The name ``humps theorem" comes from the shape of $v(x,t) \equiv v(x,0,t)$. For $t<t_0$, $v_\sigma(x,0,t)$ has two humps determined by its two symmetric maxima and its minimum for $x=x_0$. For $t\ge t_0$ all the three stationary points coincide and $v$ has a single hump.\\

\begin{figure}[!h]
\includegraphics[width=10cm]{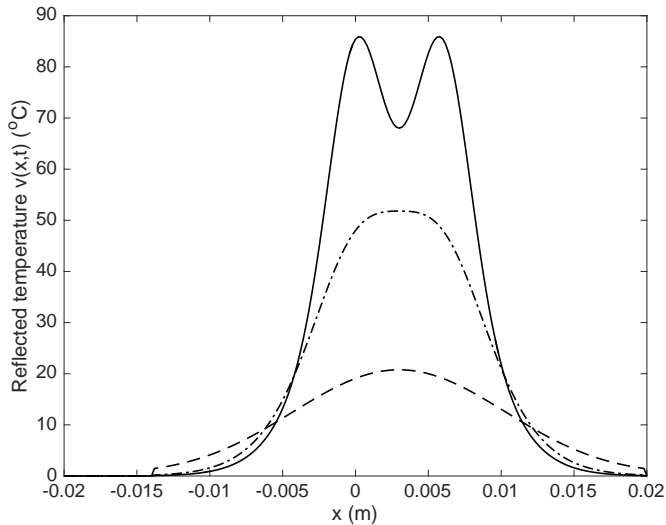}
\caption{Temperature v(x, t) reflected across the crack position, at laser OFF, time $t_L$ = 20 s (solid line), and during decay, at times 25 s (dot-dashed line) and 50 s (dashed line)}
\label{crack0}
\end{figure}

\section{Decomposition of the temperature of the slab in presence of a finite vertical crack}\label{sec:Ebi}

We have assumed that the laser spot is described by the rectangular shape defined in (\ref{power}) and it is kept ON for $t\in(0,t_{laser}<t_{max})$. Hence, 
during this lapse of time, the surface at the right hand side of the crack (more precisely the  points with $x>x_0$) does not receive any amount of  heat  through the surface $z=0$.\\

\subsection{Infinite-length crack}\label{sec:infinitecrack}

It comes from classical heat conduction theory (see \cite{CJ59} sect 14.6) that, in presence of a theoretically infinite  insulating or conducting ideal crack $\sigma=\{x_0,z\}_{0<z<b=\infty}$ (actually, a very deep one) with thermal conductance $H$,  we  have (for  $t \in (0,t_{max}]$ with $t_{max}>t_{laser}$ and $\xi>0$)
\begin{equation}\label{xminzero}
u^b(x_0-\xi,z,t)=u^0(x_0-\xi,z,t)+ u^0(x_0+\xi,z,t)-C_H(\xi,z,t)
\end{equation}
 and
\begin{equation}\label{xmagzero}
u^b(x_0+\xi,z,t)=C_H(\xi,z,t).
\end{equation}
where the function 
\begin{equation}\label{infinite}
C_H(x-x_0,z,t)=
\frac{\phi_0}{\rho c \sqrt{\pi}} \frac{H}{ \kappa} \int_{-\delta_L}^{\delta_L} dx' e^{-2 \frac{H}{\kappa} (x-x')} \int_0^t d\tau  \frac{e^{4 (\frac{H}{\kappa} )^2\alpha(t-\tau)}}{\sqrt{\alpha(t-\tau)}} f(x,x',z,t,\tau) 
\end{equation}
where
$$f(x,x',z,t,\tau) = \mathrm{erfc}\left(\frac{x-x'}{2\sqrt{\alpha(t-\tau)}}+2 \frac{H}{\kappa} \sqrt{\alpha(t-\tau)}\right) \mathrm{exp}\left(-\frac{z^2}{4\alpha (t-\tau)}\right)$$
is  smooth and positive.
In the case of insulating crack, clearly, it is $H \equiv 0$ i.e.  $C_H \equiv 0$. \\

\subsection{Finite-length crack}\label{sec:subsfem}

Assume that we are in presence of a  vertical  crack $\sigma$ of length $b>0$. The  temperature $u^b$ of our slab is the solution of the IBVP for the heat equation in $\Omega_\sigma \times (0,t_{max}]$ discussed in section \ref{sec:dama}.  We define the function
\begin{equation}\label{defE}
E(\xi,z,t)=u^b(x_0+\xi,z,t)-C_H(\xi,z,t)
\end{equation}
in $(0,\infty) \times (0,\infty) \times (0,t_{max}]$.

We have just seen that, for all $\sigma$ (Humps theorem),
\begin{equation}\label{invar}
u^b(x_0+\xi,z,t)+u^b(x_0-\xi,z,t)=u^0(x_0+\xi,z,t)+u^0(x_0-\xi,z,t)
\end{equation}
where $t \in (0,t_{\max}]$ , $\xi \in (0,\infty)$ and $x_0>0$ .\\

Hence, we have 
\begin{equation}\label{xminzeroEz}
u^b(x_0-\xi,z,t)=u^0(x_0+\xi,z,t)+u^0(x_0-\xi,z,t)-C_H(\xi,z,t)-E(\xi,z,t)
\end{equation}
for $\xi>0$.

Though the function $E$  (suitably extended to the whole real plane) looks like the temperature generated by a pair of virtual sources $S_{\pm}$ localized on the half-lines $z \ge b$ and $z \le -b$,
it can be obtained, in a more operational way, as the solution of a mixed Initial Boundary Value Problem for the heat equation. Such solution must be regarded in the weak sense \cite{Sa08} because of its non smooth behavior at the tip of the crack $P=(0,b)$. The problem of the identification of the sources $S_{\pm}$  could be studied by means of the reciprocity gap \cite{ABP13} but it will not be considered here.  \\

Thanks to the symmetries of the problem, we consider the heat equation in the first quarter $x \geq 0$ and $z \geq 0$ of $\mathbb{R}^2$ for $t \in (0,t_{max}]$:
\begin{equation}\label{heatE}
E_t(\xi,z,t)=\alpha( E_{\xi\xi}(\xi,z,t) + E_{zz}(\xi,z,t)).
\end{equation}
As for Boundary Conditions, it follows from (\ref{defE}) that 
\begin{equation}\label{Ecsi0}
-\kappa E_{\xi}(0^+,z,t)+H E(0^+,z,t)=0
\end{equation}
 for $z < b$, while
it is easy to check that 
\begin{equation}\label{bcz0}
E_{z}(\xi,0,t)=0
\end{equation}
 for all $\xi$ and $t$. 
Using (\ref{xminzeroEz}) and the continuity of $u^b$ for $z>b$, we have 
\begin{equation}
E(0,z,t)=u^0(x_0,z,t)-C_H(0,x,t)
\end{equation}
for $z > b$.  Adiabatic boundary conditions are true at infinity in the sense that $E_\xi$ and $E_z$ vanish fast for large $|z|$ and $|\xi|$. Initial data is 
\begin{equation}\label{ind0}
E(\xi,z,0)=0
\end{equation}
with $(\xi,z)\in \Omega_E$.\\

\noindent{\it Remark} The computation of the temperature in $\Omega_\sigma$ requires specific techniques in order to limit an abnormal use of memory resources and computing time due to excessively fine meshes. A recent contribution to this problem is the use of Galerkin discontinuous finite elements methods \cite{POMOCS14bis}. The decomposition $u^b=C_H+E$ developed here, reduces the computational work to the implementation of the explicit formula (\ref{infinite}) for $C_H$ plus the numerical solution of (\ref{heatE})-(\ref{ind0}) on a rectangular domain.

\section{Identification of the depth from $E$ and surface thermal measurements}\label{sec:inverse}

Numerical experiments have been conducted both on low-conductive and high-conductive materials, i.e. on materials having the thermal characteristics of concrete (thermal conductivity $\kappa\ = 1\ \mathrm{W m^{-1} K^{-1}}$, density $\rho\ = 1900\ \mathrm{kg m^{-3}}$, specific heat $c\ = 1000\ \mathrm{J kg^{-1} K^{-1}}$) and stainless steel ($\kappa\ = 14.4\ \mathrm{W m^{-1} K^{-1}}$, $\rho\ = 8000\ \mathrm{kg m^{-3}}$, $c\ = 500\ \mathrm{J kg^{-1} K^{-1}}$).

Simulations have been performed by the finite element method (FEM), using the commercial code COMSOL Multiphysics$\textsuperscript{\textregistered}$ \cite{Comsol}, on domains containing cracks of realistic shapes and length. The purpose is that of obtaining the ``vertical-equivalent'' crack depth by means of the model described in Section \ref{sec:subsfem}.

\subsection{Simulation involving a concrete wall}
Figure \ref{crack0} shows a crack, starting vertically from the surface and developing inside concrete up to a depth of $8$ mm. A laser heats the specimen two millimeters to the left of the crack opening. The spot is the segment $(-\epsilon,\epsilon)$ with $\epsilon = 1$ mm (since the simulation is 2D,  this spot is equivalent to  a line-source of width $2$ mm). The assumed power density is $\mathrm{4 \times 10^4\ W/m^3}$. Laser power is switched off after $40$ seconds. 
The wall thickness ($4$ cm)  is virtually infinite for the time interval ($t \le 600$ s)  of interest of the simulation as observed in  section \ref{sec:diffusivity}. The crack width is $0.2$ mm.

\begin{figure}[!h]
\begin{center}
\includegraphics[width=10cm]{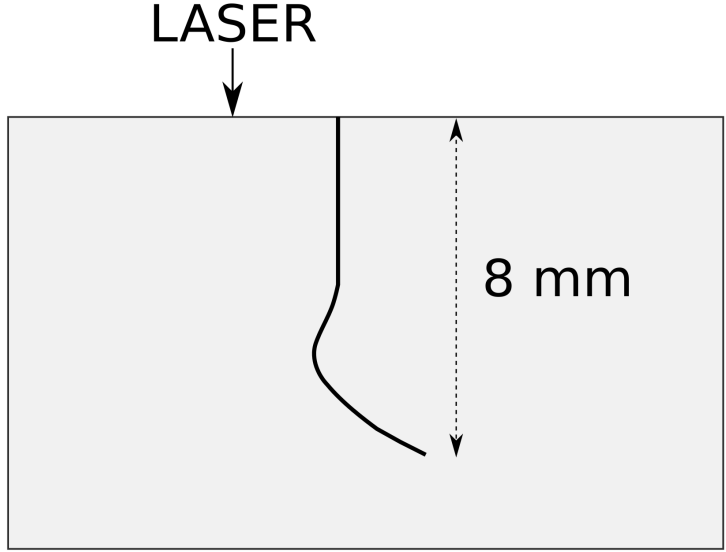}
\caption{Realistic shape and size of a crack in a concrete slab}
\label{crack0}
\end{center}
\end{figure}

\begin{figure}[!h]
\begin{center}
\includegraphics[width=10cm]{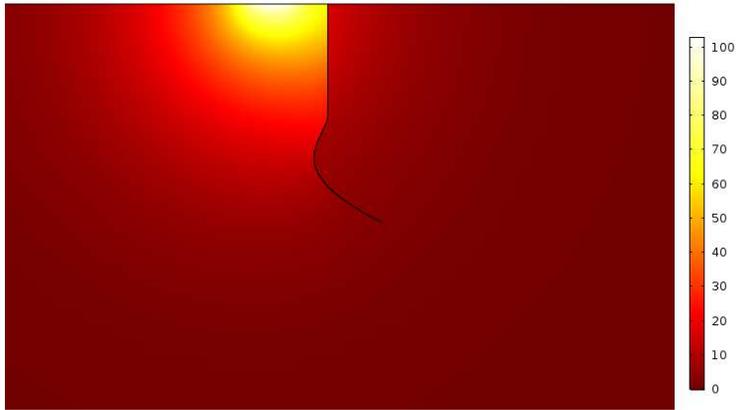}
\caption{Temperature at laser off time for left-side illumination}
\label{crack1}
\end{center}
\end{figure}

\begin{figure}[!h]
\begin{center}
\includegraphics[width=10cm]{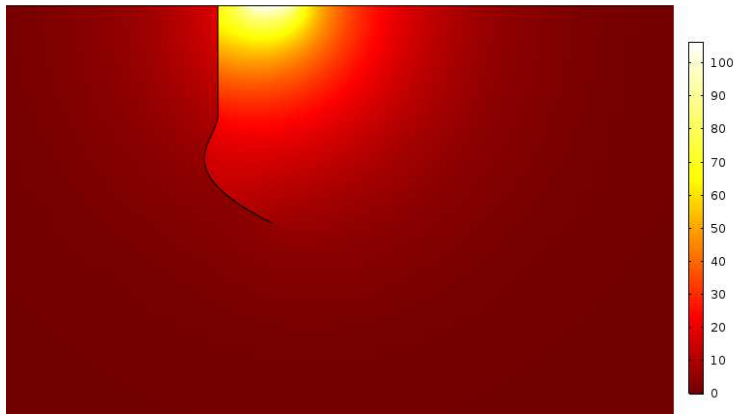}
\caption{Temperature at laser off time for right-side illumination}
\label{crack2}
\end{center}
\end{figure}

Figure \ref{crack1} and \ref{crack2} show the temperature distributions (in $\mathrm{^o C}$) at the end of laser heating (i.e. at $t = 40$ s), for laser illumination respectively located 2mm to the left and to the right of the crack.

The inversion procedure minimizes the discrepancy $Err(t)$ between the measured temperature $u_{meas}^b$ and  computations of $u_{comp}^b$  carried out following the decomposition introduced in section \ref{sec:Ebi}. The function  $Err$ is defined as:
\begin{equation}
Err(t) = \sum_{i=1}^N \left( u_{meas}^b(x_0+\xi_i,0,t) - u_{comp}^b(x_0+\xi_i,0,t) \right)^2 
\end{equation}
where $x_0+\xi_i$ ($i=1,...,N)$ are points of the positive $x$-axis taken on the right of the crack opening ($x_0$), at a suitable time $t$. 

\begin{figure}[!h]
\begin{center}
\includegraphics[width=10cm]{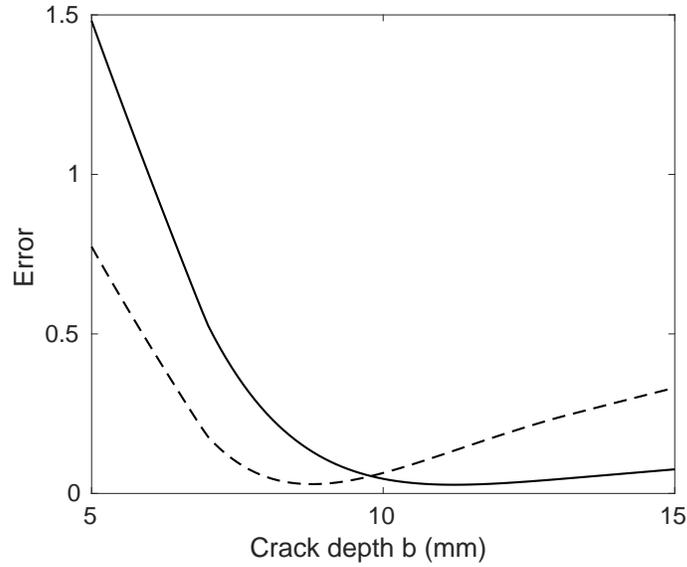}
\caption{Error at time t = 200 s for left-side (solid line) and right-side (dashed line) illumination}
\label{err_cemento}
\end{center}
\end{figure}

Figure \ref{err_cemento} shows the error at time t = 200 s, for laser illumination respectively located 2mm to the left (solid line) and to the right (dashed line) of the crack. The minimum is located at about 11 mm for left illumination and at 8 mm for right illumination. The former is close to the actual crack length, while the latter is exactly the vertical depth of the crack.

\subsection{Simulations involving a stainless steel slab}
We have performed two simulations on a steel slab of $2$ cm thickness, with specular illuminations (see Figure \ref{crack1}).
The crack starts vertically in $x=x_0$, changes its direction around its half depth and develops inside the metal up to a depth of $7$ mm. The laser spot is positioned in $x=0$ and $x=2x_0$, i.e. two millimeters to the left or to the right of the crack opening. 

Power width is $2$ mm (again, the simulation is 2D so we are dealing with line-sources) and the assumed power density is $\mathrm{4 \times 10^5\ W/m^3}$. Laser power is switched off after a time of $40$ seconds. 
The crack width is $0.1$ mm.

\begin{figure}[!h]
\begin{center}
\includegraphics[width=10cm]{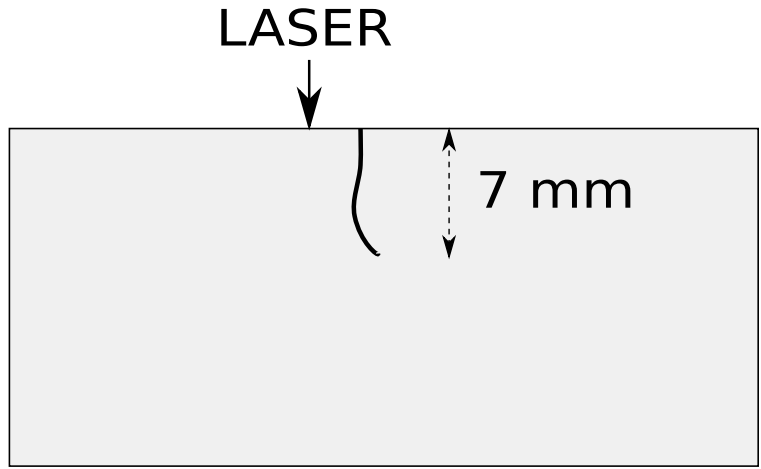}
\caption{Realistic shape and size of a crack in a stainless steel slab}
\label{crack_1}
\end{center}
\end{figure}

\begin{figure}[!h]
\begin{center}
\includegraphics[width=10cm]{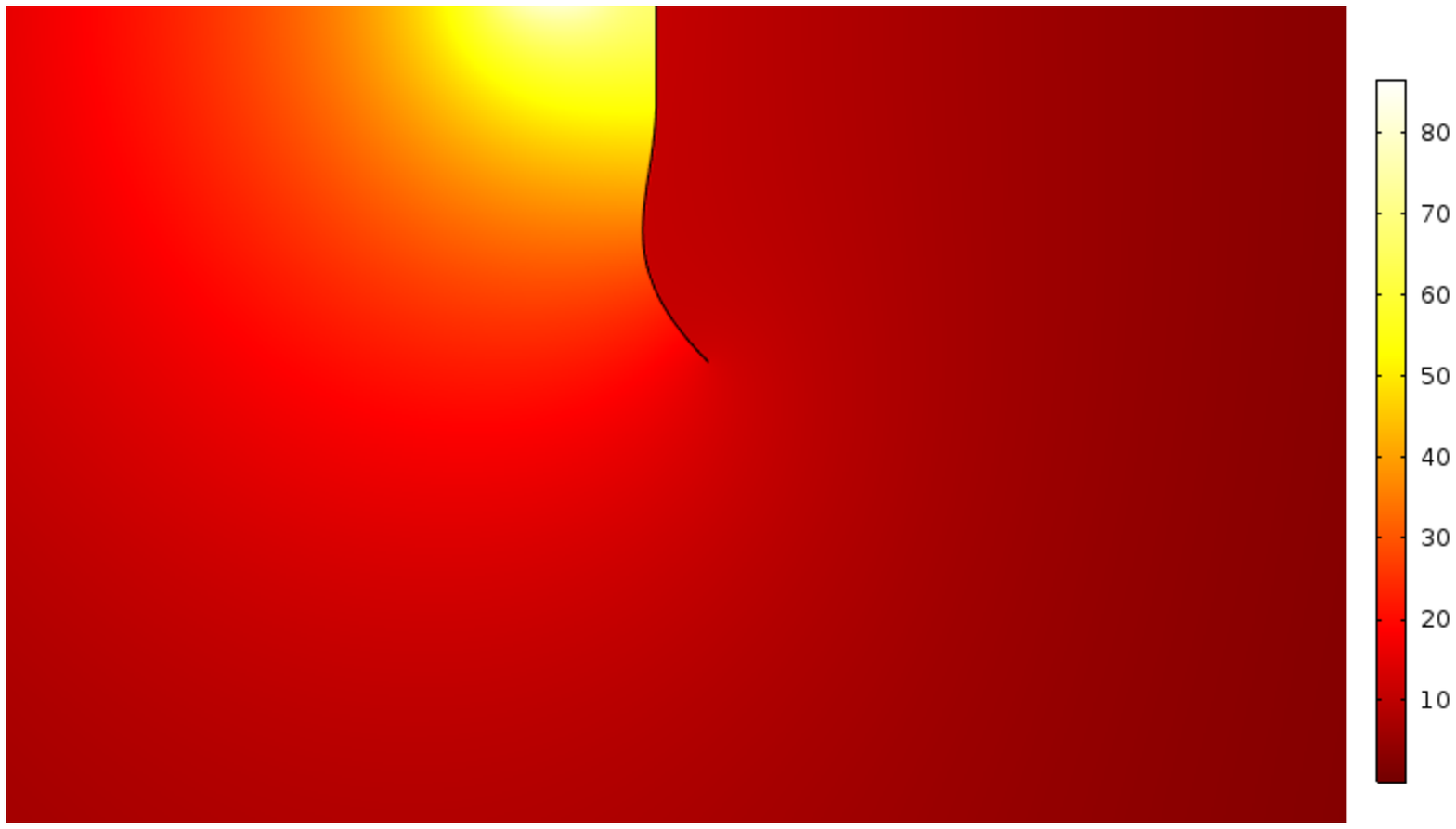}
\caption{Temperature at laser off time for left-side illumination}
\label{crack1_T}
\end{center}
\end{figure}

\begin{figure}[!h]
\begin{center}
\includegraphics[width=10cm]{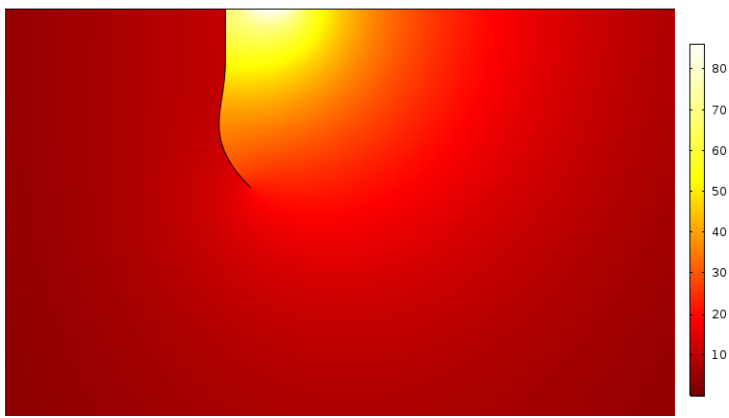}
\caption{Temperature at laser off time for right-side illumination}
\label{crack2_T}
\end{center}
\end{figure}

Figures \ref{crack1_T} and \ref{crack2_T}
show the temperature distribution (in $\mathrm{^o C}$) at the end of laser heating (i.e. at $t = 40$ s) when the laser is respectively pointed to the left and to the right of the crack opening.

\begin{figure}[!h]
\begin{center}
\includegraphics[width=10cm]{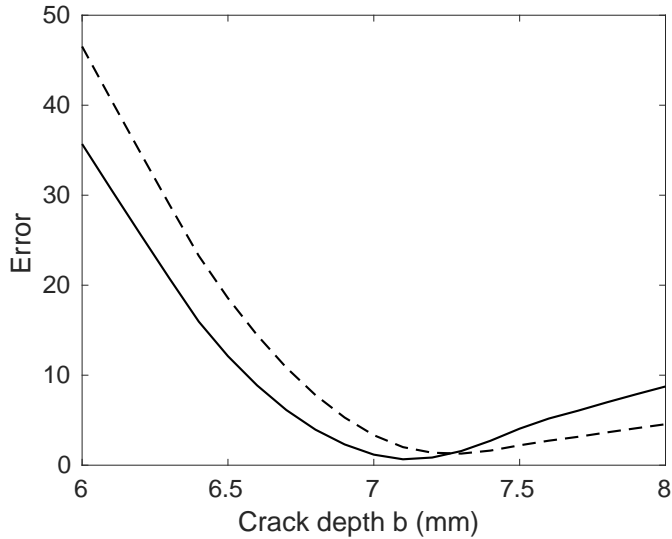}
\caption{Error at time t = 40 s for left-side (solid line) and right-side (dashed line) illumination}
\label{err_metal}
\end{center}
\end{figure}

Figure \ref{err_metal} compares the function $Err(t)$  at time $t = 50$ s for the two cracks (first crack solid line, second crack dashed line).

\subsection{Conclusions}  Nondestructive evaluation of  crack depth from Laser Spot Thermography (spot centered in $(0,0)$ and laser on for$t \le t_{laser}$), requires the numerical solution of an inverse problem by means of an iterative method. Hence, the core of any inversion strategy is the solution of the underlying IBVP for the heat equation.
Finite element solution of the heat equation in a fractured domain $\Omega \setminus \sigma$ demands for specific techniques to handle  extremely fine meshes. In this paper we have developed a method for studying this IBVP in a simpler way. First, we introduce the temperature $C_H(\xi,z,t)$ corresponding to the case in which an infinite vertical crack of thermal contact coefficient $H$ splits the specimen. The analytical form of $C_H$ is known (see section \ref{sec:infinitecrack}). Than, we define a function $E_b$ that gives a measure of the temperature  due to the heat passing under the crack and such that $C_H+E_b$ is the temperature of the specimen on the right hand side of the crack. We show that, if $\sigma$ is a segment parallel to the $z$-axis,  $E_b$ solves a simple IBVP for the heat equation in a quarter of plane. In this way we can implement an inexpensive numerical method for the minimization, varying the parameter $b$, of the discrepancy between $u^b_{computed}=(C_H+E_b)(\xi,0,t_i)$ and the sequence of thermal maps $ u^b_{measured}(x_0+\xi,0,t_i$ of the surface $z=0$ for $t_1<t_2<...<t_N<\tau_ch$. The value of $b$ corresponding to the minimum is an estimate of the depth of the unknown crack. It is remarkable that in this way we determine the depth of a virtual vertical crack like $\sigma$, but, as shown in simulations, we have also an approximate evaluation of the length of a perturbation of $\sigma$.

\subsection{Acknowledgements} 
This work has been supported by the  FAR-FAS research project ``TOSCA-FI'' funded by Tuscany Region.

\end{document}